\documentclass[fleqn,usenatbib]{mnras}

\usepackage{newtxtext,newtxmath}
\usepackage[T1]{fontenc}

\DeclareRobustCommand{\VAN}[3]{#2}
\let\VANthebibliography\thebibliography
\def\thebibliography{\DeclareRobustCommand{\VAN}[3]{##3}\VANthebibliography}

\usepackage{graphicx}	% Including figure files
\usepackage{amsmath}	% Advanced maths commands
\usepackage{caption}
\usepackage{subcaption}
\usepackage{graphicx}
\title[Absorption of GRBs by AGN]{The effects of Time-Variable Absorption due to Gamma-Ray Bursts In Active Galactic Nuclei Accretion Disks}

\author[Michael Ray]{
Michael Ray$^{1}$\thanks{E-mail: michael.ray436@gmail.com},
Davide Lazzati$^{2}$,
Rosalba Perna$^{1,3}$
\\
$^{1}$Department of Physics and Astronomy, Stony Brook University, Stony Brook, NY 11794-3800, USA\\
$^{2}$Department of Physics, Oregon State University, 301 Weniger Hall, Corvallis, OR 97331, USA\\
$^{3}$Center for Computational Astrophysics, Flatiron Institute, New York, NY 10010, USA
}

\date{}

\pubyear{2022}

\begin{document}
\label{firstpage}
\pagerange{\pageref{firstpage}--\pageref{lastpage}}
\maketitle

% Abstract of the paper
\begin{abstract}
Both long and short gamma-ray bursts (GRBs) are expected to occur in the dense environments of active galactic nuclei (AGN) accretion disks. As these bursts propagate through the disks they live in, they photoionize the medium causing time-dependent opacity that results in transients with unique spectral evolution. In this paper we use a line-of-sight radiation transfer code coupling metal and dust evolution to simulate the time-dependent absorption that occurs in the case of both long and short GRBs. Through these simulations, we investigate the parameter space in which dense environments leave a potentially observable imprint on the bursts. Our numerical investigation reveals that time dependent spectral evolution is expected for central supermassive black hole masses between $10^5$ and $5 \times 10^7$ solar masses in the case of long GRBs, and between $10^4$ and $10^7$ solar masses in the case of short GRBs. Our findings can lead to the identification of bursts exploding in AGN disk environments through their unique spectral evolution coupled with a central location. In addition, the study of the time-dependent evolution would allow for studying the disk structure, once the identification with an AGN has been established. Finally, our findings lead to insight into whether GRBs contribute to the AGN emission, and which kind, thus helping to answer the question of whether GRBs can be the cause of some of the as-of-yet unexplained AGN time variability.

\end{abstract}

% Select between one and six entries from the list of approved keywords.
% Don't make up new ones.

%%%%%%%%%%%%%%%%%%%%%%%%%%%%%%%%%%%%%%%%%%%%%%%%%%

%%%%%%%%%%%%%%%%% BODY OF PAPER %%%%%%%%%%%%%%%%%%

\section{Introduction}

Gamma-ray bursts (GRBs) are among the most energetic events in the Universe, capable of producing peak observed bolometric luminosities greater than $10^{53}$ erg s$^{-1}$ \citep{Gehrels2009}. They come in two varieties, long and short, which are distinguished based on the time-scale in which their prompt emission (early time $\gamma$-ray emission) is observed \citep{Kouveliotou1993}. Short GRBs are commonly defined to be those whose prompt phase lasts for two seconds or less and are believed to result from compact object mergers (\citealt{Fong2013, Belczynski2006, Mochkovitch1993};  at least one short GRB has already been confirmed to be the result of neutron star mergers, \citealt{Ligo-Virgo2017A, Ligo-Virgo2017B}), while long GRBs are those which last longer than two seconds and are believed to result from the collapse of massive stars \citep{Hjorth2003,Stanek2003, McFayden1999, Heger2003}. Because of their enormous energy output, both long and short GRBs can be seen from across the observable Universe, making them ideal sources to use to study distant galaxies.

The prompt emission of GRBs tends to have complex time variability and a spectrum that is generally described by a  broken power law with just three parameters \citep{Band1993}, while the spectrum of the afterglow emission (late-time radiation) is a power-law with multiple breaks due to injection, cooling, and absorption \citep{Sari1998}.

Due to the simplicity of their afterglow spectra, GRBs are also ideal candidates to probe  the medium in which they are emitted by observing the absorption lines imprinted on their spectra. While time-dependent absorption of GRB spectra in various media has been studied extensively (see e.g. \citealt{Perna1998,Bottcher1998,Lazzati2001, Frontera2004,Robinson2009, Campana2021}), bursts in the environment of Active Galactic Nuclei (AGN) accretion disks is a relatively new area of research with few dedicated studies thus far (for studies that have already been performed, see e.g. \citealt{Perna2021-a, Yuan2021, Zhu2021-a, Zhu2021-b,Lazzati2022}).

Active Galactic Nuclei (AGNs) are galactic centers with much higher than normal luminosity that is not characteristic of stellar emission. The emission from AGNs is believed to be driven by an accretion disk powering a central supermassive black hole (SMBH; \citealt{Woo2002}). While this accretion process is well understood, there is notable time variability observed in AGN spectra that has yet to be fully explained \citep{BPeterson2001}. Some have suggested that this variability is the result of stochastic temperature fluctuations in the accretion disk, modelled by a damped random walk \citep{Kelly2009,MacLeod2010, Zeljko2014,Kozlowski2016}. Others have cast doubt on whether this is a viable model of AGN variability \citep{Zu2013, Mushotzky2011,Kasliwal2015}. An alternative and perhaps complementary explanation  is that AGN variability, or at least a fraction of it, is caused by GRBs or other stellar transients emitted from within the AGN accretion disk. This possibility is made more plausible by the observation that AGN accretion disks are dense environments that carry stars as a result of both in-situ formation
\citep[e.g.][]{Paczynski1978,Goodman2003,Dittmann2020},  and capture from the nuclear star cluster surrounding the AGN  \citep[e.g.][]{Artymowicz1993,Fabj2020} due to  momentum and energy loss as the stars interact with the disk. Evolution in AGN disks results not only in mass growth \citep{Cantiello2021,Dittmann2021,Dittmann2022}, but also angular momentum growth, which makes AGN stars likely to end their lives with the right conditions to produce a GRB \citep{Jermyn2021}.
Additionally, frequent dynamical interactions 
within AGN disks (e.g. \citealt{Tagawa2020}) result in frequent binary formation, and hence the potential to yield short GRBs when two neutron stars, or a neutron star and a black hole, merge.
While we constrain ourselves to GRBs in this paper, AGN disks are also expected to host various events capable of producing electromagnetic transients such as tidal disruption events \citep{Yang2022}, accretion-induced collapse of neutron stars \citep{Perna2021-b}, core-collapse supernovae \citep{Grishin2021, Cantiello2021}, and binary black hole mergers \citep{Graham2020, Grobner2020}.

In this paper, we study the absorption that a dense environment causes on long (LGRB) and short GRB (SGRB) spectra, predominantly due to its photoelectric absorption. We then specialize it to the environment of an AGN disk, described by the disk models of \citet{Sirko2003} and \citet{Thompson2005}.
We perform a grid of simulations %\textbf{under the assumption of a geometrically thin accretion disk model,} 
to identify the conditions under which dense environments have a sizable and time-variable effect on GRB spectra (the precise meaning of "sizable and time-variable effect" is given in section \ref{sec:Simulation_Description_and_Results}). More specifically, since the early, high energy radiation from the GRB photoionizes the medium, it results in a time-dependent opacity during the early life of the transient. Since the medium opacity affects spectra from the X-rays through the optical band, 
the combination of the intrinsic GRB spectrum with a variable opacity can produce unusual transients with a recognizable spectral evolution. To study and quantify this effect,
we use a radiation transfer code developed by \citet{Perna2002}, allowing us to calculate the effects that the dense environment induces on the GRB spectrum.

This study is organized as follows: in \S 2 we present the setup of the simulations including a description of how the radiation transfer code works, a description of the GRB luminosity functions used, as well as a description of the properties of the absorbing medium. In \S 3 we present a detailed description of the simulations performed and the parameter space that is covered in terms of medium properties. 
We then present the results of these simulations. In \S 4 we discuss the conclusions that can be drawn from the study and we also comment on future work to be done to extend and generalize the findings of this study.

\section{Simulation Setup}

\subsection{Choice of central densities and density profiles}
The radiative transfer code used \citep{Perna2002} is flexible to any density and temperature profile desired. Here, our goal is to measure where in the ($n_0$, $H$) parameter space the effect of absorption variability is potentially observable, where $n_0$ represents the density of neutral atomic Hydrogen and $H$ represents the AGN scale height. To find this region in the parameter space, we perform a grid of simulations over a wide range of combinations of $n_0$ and $H$.
We exclude any combinations of $n_0$ and $H$ where the absorbing medium column density, $N_{\rm H} = n_0 H$, is greater than $10^{24}$ cm$^{-2}$. This is because at column densities greater than $\sim10^{24}$ cm$^{-2}$ the medium becomes optically thick to Thompson scattering. In these conditions, all photons interact with the medium either by being photoabsorbed or by being Thomson scattered. Even though a scattered photon does not disappear, its path to the observer is increased by $\sim \tau_T H$, where $\tau_T$ is the Thomson opacity of the medium. This causes diffusion of the prompt emission over a timescale 
{$ \Delta{t}\sim 3 (H/10^{16}~{\rm cm})^2(n_0/10^9~{\rm cm}^{-3})\tau_T$~{ days}}. Such a temporally stretched burst would be undetectable in most circumstances (see also \citealt{Wang2022} for a more refined treatment of radiative transfer under these conditions).

We assume the density of the medium along the z-axis (taking z to be the line-of-sight coordinate) to be an (unnormalized) Gaussian centered around $z=0$ and with standard deviation $H$,
\begin{equation}
    \rho(z) = \rho_0 \textrm{ exp}\left(\frac{-z^2}{2H^2}\right)\,,
\end{equation}
\noindent
where $\rho_0 = n_0 m_p$ is the mass density at the center of the disk.
%REMOVED the few sentences below b/c it's a duplicate of the paragraph at the very end of section 3.1
%One should take note that our results are not necessarily particular to AGN disks. In fact, this analysis will apply to GRBs that are emitted in any location where the density along the line-of-sight falls off as a Gaussian with scale height  $H$. Our results should also be broadly applicable to any density profile characterized by a sharp decline at the outer edge, as is the case in AGN disks.

\subsection{GRB Light Curves and Spectra}
We model the GRB luminosity as the sum of a prompt emission component and an afterglow component such that the total light curve is given by
\begin{equation}
    L(t,\nu) = L_{p}(t,\nu) + L_{AG}(t,\nu)\,,
    \label{eqn:lum_sum}
\end{equation}

\noindent 
where $L_p(t, \nu)$ is the light curve of the prompt emission and $L_{AG}(t, \nu)$ is the light curve of the afterglow.
For the prompt emission, the luminosity separates into a time-dependent component and a frequency-dependent component, that is,
\begin{equation}
    L_p(t,\nu) = A_p T_p(t) F_p(\nu)\,,
\end{equation}
\noindent
where $A_p$ is a normalization constant \citep{Robinson2009}. For the afterglow emission, we use a code that numerically computes afterglow emission in a given environment and we then fit analytical luminosity curves to the output of the computation.

\subsubsection{Long GRBs - Prompt Emission}\label{sec:LGRBPE_Luminosity}
We model the LGRB prompt emission following the analytical fits derived by \citet{Robinson2009}. In their model, the functions $T_p(t)$ and $F_p(\nu)$ are each independently normalized such that $\int_0^{\infty} F_p(\nu) d\nu=1$ and $\int_{0}^{\infty} T_p(t) dt=1$. This ensures that the constant $A_p$ contains all of the normalization for the prompt emission. The time-dependent component of the prompt emission for the LGRB takes the form of a Gaussian with mean $10$ seconds and full width half max $10$ seconds. Thus,

\begin{equation}
    T_p(t) = \frac{A_{pt}}{\sqrt{2\pi\sigma^2}} e^{-\frac{(t-t_0)^2}{2\sigma^2}}\,,
    \label{eqn:T_p(t)}
\end{equation}

\noindent
where $t$ is measured in seconds, $t_0=10$~s, and $\sigma=10/(2\sqrt{2\log2)}\simeq4.25$~s. Normalizing such that $\int_0^{\infty} T_{p}(t) dt = 1$ gives $A_{pt} = 1.00935$.

The frequency-dependent component of the LGRB prompt emission takes after the spectrum given in \citet{Band1993} and is modeled by a broken power-law as:

\begin{equation}
    F_{p}(E) = A_{pf} \left(\frac{E}{100 \textrm{ keV}}\right)^\alpha \textrm{ exp}\left(- \frac{E}{E_0}\right),
    \;\; (\alpha - \beta)E_0 \geq E
    \label{eqn:lowE_flux_LGRB}
\end{equation}
\begin{multline}
    F_{p}(E) = A_{pf} \left[\frac{(\alpha - \beta)E_0}{100 \textrm{ keV}}\right]^{\alpha-\beta} \textrm{ exp}(\beta - \alpha) \left(\frac{E}{100 \textrm{ keV}}\right)^{\beta},\\
     (\alpha - \beta)E_0 < E
    \label{eqn:highE_flux_LGRB}
\end{multline}
  
\noindent 
where ($\alpha$ - $\beta$) $E_0$ is the knee of the power-law taken to be 300 keV (i.e. this is the energy at which the luminosity function "turns over"), $\alpha$ is the slope of $F_{p}(E)$ for $E \leq E_0$ and $\beta$ is the slope of $F_{p}(E)$ for $E > E_0$, and $A_{pf}$ is a normalization constant. We use $\alpha=0$ and $\beta=-2$ in our models. For these values of $\alpha$ and $\beta$ it is easy to solve the equation $\int_0^{\infty} F_p(E) dE = 1$ for $A_{pf}$. Doing this leads to $A_{pf} = 2.935990 \times 10^{-3}$ keV$^{-1}$. Thus, we arrive at the prompt emission light curve given by:

\begin{equation}
    L_{p} = A_{p} T_{p}(t) F_{p}(E)
\end{equation}
  
\noindent 
where $T_{p}(t)$ and $F_{p}(E)$ are defined above. The constant $A_{p}$ is the total energy output of the prompt emission, given by $A_{p} = \int_{t=0}^{\infty} \int_{\nu=0}^{\infty} L_p(t,\nu) d\nu dt$. Here we take $A_p = 10^{53}$ ergs.

\begin{figure}
     \centering
     \begin{subfigure}[b]{\columnwidth}
         \centering
         \includegraphics[width=\columnwidth]{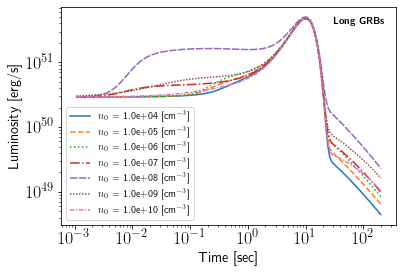}
     \end{subfigure}
     \hfill
     \begin{subfigure}[b]{\columnwidth}
         \centering
         \includegraphics[width=\columnwidth]{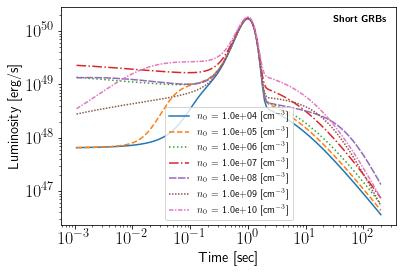}
     \end{subfigure}
     \caption{Input luminosity as a function of time in the 15keV to 150keV band for both long (top) and short (bottom) GRBs. 
     The functional form for the prompt emission is the same in both cases (but with different parameters)
     and calculated analytically as presented in \S \ref{sec:LGRBPE_Luminosity} and \ref{sec:SGRBPE_Luminosity}, while the afterglow emission is numerically
     computed as described in \S \ref{sec:AG_Luminosity}.
     } 
    \label{fig:input_lum_curves}
\end{figure}

\subsubsection{Short GRBs - Prompt Emission}\label{sec:SGRBPE_Luminosity}
For our model of a short GRB (SGRB), we use a similar method as with the LGRB, using Eq.~\ref{eqn:lum_sum} to split the luminosity into a prompt emission component and an afterglow component. The goal of our analysis is to provide an example of absorption of a "typical" SGRB. While GRBs do vary quite a lot in their exact time-dependent spectrum, we use the average properties of many SGRBs to create a model of "typical" SGRB prompt emission. The time-dependent component of the SGRB prompt emission takes the same form as Eq.\ref{eqn:T_p(t)} with new parameters $\sigma = 0.3$ s, $t_0 = 1$ s, and $A_{pt} = 1.00043$.

For the frequency-dependent component of the SGRB prompt emission, we take a functional form identical to that of the LGRB, given in Eqs.\ref{eqn:lowE_flux_LGRB}, \ref{eqn:highE_flux_LGRB}. For the parameters here we use those that are typical for SGRBs: $\alpha = 0.5$, $\beta = -1.5$, and $E_0 (\alpha - \beta) = 350$ keV \citep{Ghirlanda2009}. Then just as before, we take $\int_{E=0}^{\infty} F_{p}(E) dE = 1$, giving $A_{pf} = 1.68928 \times 10^{-3}$ keV$^{-1}$. Thus, the total prompt emission lightcurve for the SGRB is given by:

\begin{equation}
    L_p = A_p T_p(t) F_p(E)\,,
\end{equation}

\noindent
where $T_p(t)$ and $F_p(E)$ are defined in the preceding paragraph. The constant, $A_p = \int_{t=0}^{\infty} \int_{\nu=0}^{\infty} L_p(t, \nu) d\nu dt$, is the total energy output of the SGRB prompt emission and is taken to be $10^{51}$ ergs, as is typical for a SGRB \citep{Fong2015}.

\subsubsection{Afterglow Emission}\label{sec:AG_Luminosity}
While GRB afterglows are still a current topic of active research (see e.g. \citealt{Golant2022}), it is currently understood that afterglows for both short and long GRBs are synchrotron radiation resulting from the collision of a relativistic shell with an external medium \citep{Sari1998, Panaitescu2000}. 
To compute our afterglows, we used a code that has been used in various previous papers \citep{Lazzati2018, Perna2022, Wang2022} and performs the afterglow computation numerically. As input parameters to the afterglow computation we use $\epsilon_e = 0.3$ and $\epsilon_B = 0.1$, where $\epsilon_e$ is the fraction of the shock energy that is given to electrons and $\epsilon_B$ is the fraction of the shock energy given to tangled magnetic fields. We take the electron acceleration to have a distribution given by $n(\gamma) \propto \gamma^{-p_{el}}$ with $p_{el} = 2.3$. Finally, we take a uniform ISM medium surrounding the burst, which is a good approximation to the true matter distribution because the burst is at the center of the AGN disk where the Gaussian matter distribution is nearly constant.

To optimize our code run-time, we then fit analytical curves to the numerically computed curves. The fit that we use for each of the curves assumes a broken power-law shape in both time and frequency \citep{Sari1998, Panaitescu2000, Granot2002, Rossi2022}.
Since the shell/medium collision dynamics will depend on the density of the medium in the immediate vicinity of the burst, each afterglow model with a different value of $n_0$ will have different parameters. 

The full input luminosity curves (that is, the sum of prompt emission and afterglow emission) in the 15keV to 150keV band are shown in Figure \ref{fig:input_lum_curves} as a function of time. 
The value of the scale height does not affect the input GRB spectrum and thus does not enter into this calculation. The scale height will, however, affect the absorption of the input spectrum, and thus will affect the output spectrum observed.

\subsection{Numerical Setup and Code Description}
At the core of the simulations performed is a radiation transfer code which takes into account the time-dependent photo-ionization of both dust and metals in a medium subjected to an intense radiation field \citep{Perna2002}. The code computes, on a 2-d space-time grid (one line-of-sight spatial coordinate and one time coordinate), the state of the radiation field, the abundance and ionization states of both molecular and atomic Hydrogen, and the abundance and ionization states of the 12 next most common astrophysical elements: He, C, N, O, Ne, Mg, Si, S, Ar, Ca, Fe, Ni. In addition to computing the state of the medium and radiation field at each grid point, the code calculates the output flux spectrum (that is, the flux emanating from the outermost bin along the z-axis, located at $z_{max}$) and the (frequency-dependent) optical depth, both as functions of time. In this way, the code produces  a time-dependent optical depth spectrum and a time-dependent flux spectrum which fully describes the radiation that emerges from the dense environment and flows freely to an observer. The radiative transfer is calculated in the energy range of $1$ eV to $50$ keV and throughout the calculations, energies are binned into 200 equally spaced bins.

When setting up the space-time grid, we take the start time, $t_i$, to be $10^{-3}$ seconds and the end time, $t_f$, to be $200$ seconds with $1500$ logarithmically spaced time steps. We take the minimum z-coordinate to be $z_{min} = 1.25 \times 10^{-3} H$ where $H$ is the scale height of the AGN and the constant in front is chosen such that 0.1\% of the total mass is contained within $z_{min}$. The maximum z-coordinate is taken to be $z_{max} = 2.58 H$ where the constant is chosen such that 99\% of the total mass is contained within $z_{max}$. The interval $[z_{min}, z_{max}]$ is split into 100 logarithmically spaced steps.

We take the initial temperature to be $10^4$ K in all simulations regardless of the values of $n_0$ and $H$. The constant temperature choice is motivated by the fact that the radiative transfer is much more sensitive to the density of the medium than the temperature of the medium. This phenomenon can be understood by noting that, for a range of initial medium temperatures ($\sim 10^2-10^4$~K), over a short period of time (i.e. the recombination time) the medium will be heated by the X-ray/UV early afterglow radiation to a value that depends largely on the photo-ionizing spectrum, and hence our results are not very sensitive to the initial temperature of the medium.

\section{Simulation Description and Results}
\label{sec:Simulation_Description_and_Results}
% Our goal is to identify the area of the $(n_0, H)$ parameter space in which we get \textbf{"interesting"} absorption of our bursts. What remains to be defined, however, is what we mean by \textbf{"interesting absorption". What we want to look for is bursts that are absorbed by the medium to a significant level (i.e. the burst does not simply freely flow from the source to the observer), while not being completely absorbed by the medium. Additionally, in our definition, we want to impose that there is significant time variability in the absorption, such that we ensure there is feedback between the radiation and the medium throughout the duration of the simulation. To achieve these goals, we propose the following three rules to determine whether "interesting absorption" has occurred:}

Our goal is to identify the area of the $(n_0, H)$ parameter space in which gamma-ray bursts would be characterized by potentially observable variable absorption.  We consider that variable absorption is potentially observable
if the absorption along the line of sight meets two conditions: i) it is initially substantial for at least a fraction of a second 
%ii) it does not vanish on a time scale that is too short to be observed; 
and ii) it changes by at least a factor 2 during the 
duration of the simulation. We quantify these requests by imposing:

\begin{itemize}
    %\item $\tau_{0.1-10}(t_f) < 1$
    \item[\bf{1.}] $\tau_{0.1-10}(0.1\textrm{ sec}) > 0.7$
    \item[\bf{2.}] $\tau_{0.1-10}(0.1\textrm{ sec}) / \tau_{.1-10}(t_f) > 2$\,.
\end{itemize}

\noindent
In the above, $\tau_{0.1-10}(t)$ is defined as the average optical depth between energies 0.1 keV and 10 keV at time $t$, and $t_f$ is the maximum time of the simulation, taken to be $200$ seconds. 

These conditions are somewhat arbitrary and only a detailed analysis on a burst-by-burst case would truly establish whether or not variable absorption can be detected. However, if these conditions are met, variable absorption should be detectable even for bursts that are not bright enough to allow for a detailed time-resolved spectroscopic study. Hence they can be considered as conservative conditions.
%These conditions ensure three things in the simulations:

% \begin{itemize}
%     \item[1.] $\tau_{0.1-10}(t_f) < 1$ \textbf{ensures that at time $t_f$ (at the end of the simulation) the flux flowing out of the medium is no less than $\frac{\textrm{flux emitted by GRB}}{e}$. While the exact choice of the requisite ratio of outgoing flux to input flux is relatively arbitrary, our choice here guarantees we do not label overly or entirely absorbed bursts as "interesting"}.
%     \item[2.] $\tau_{0.1-10}(0.1 \textrm{ sec}) > 0.7$ \textbf{ensures that at time $\mathbf{t=0.1}$ sec, the flux flowing out of the surface of the medium is no greater than $\frac{\textrm{flux emitted by GRB}}{e^{0.7}} \approx \frac{\textrm{flux emitted by GRB}}{2}$. This guarantees that we have some significant absorption happening at the beginning of the simulation (i.e. the emitted radiation is not simply moving unobstructed through the medium).}
%     \item[3.] $\tau_{0.1-10}(0.1\textrm{ sec}) / \tau_{.1-10}(t_f) > 2$ ensures that there is some significant change in the optical depth over the course of the simulation. This means that there is some dynamical feedback between the radiation and the medium over the course of the simulation time.
% \end{itemize}

% Now that we have defined what we are searching for in our simulations, we can present our findings and determine which combinations of $n_0$ and $H$ provide an environment where we observe \textbf{interesting} absorption that also varies in time over the course of the simulation.

\subsection{Burst Classification}
\label{sec:Burst_Classification}

\begin{figure}
     \centering
     \begin{subfigure}[b]{\columnwidth}
         \centering
         \includegraphics[width=\columnwidth]{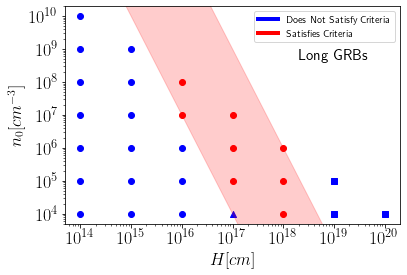}
     \end{subfigure}
     \hfill
     \begin{subfigure}[b]{\columnwidth}
         \centering
         \includegraphics[width=\columnwidth]{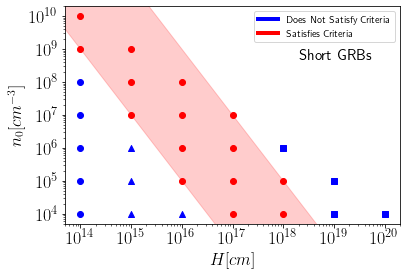}
     \end{subfigure}
     \caption{A map showing the parameter space for both long (top) and short (bottom) GRBs where the criteria is satisfied for "potentially observable variable absorption", as defined in section \ref{sec:Simulation_Description_and_Results}. Red dots indicate the simulation satisfied all of our criteria for potentially observable variable absorption, while blue markers indicate the opposite. The meaning of different shapes is indicated in table \ref{table:marker_meanings}. The red shaded areas indicate the bounds laid out in Eqns.~\ref{eqn:bounds_on_interesting_LGRBs}, \ref{eqn:bounds_on_interesting_SGRBs}.
     }
     \label{fig:Significant_absorption_map}
\end{figure}

\maketitle % <==========================================================
\begin{table*} % <======================================================
  \caption{Definition of marker shapes for all figures that classify bursts based on their absorption variability properties and potential for detection}\label{table:marker_meanings} % <========================
\begin{tabular}{ |p{4cm}|p{12cm}| }
 \hline
 \hline
 \textbf{Marker Shape} & \textbf{Marker Meaning}\\
 \hline
 Red Circle & Both criteria one and two are satisfied. The absorption variability is potentially observable.
 \\
 \\
 
% Blue Circle & The only criteria that is met is the criteria on $\tau_{0.1-10}(t_f)$. This indicates that there was too little absorption for the entire duration of the simulation, including at early times.
Blue Circle & The simulation fails both criteria one and two. This indicates that the absorption is neither initially substantial nor highly variable in time.
 \\
 \\

 Blue Square & Criterion one is satisfied, but not criterion two. This indicates that the absorption is initially substantial, but is not highly variable throughout the time of the simulation.
 \\
 \\
 
 Blue Triangle & Criterion two is satisfied, but criterion one is not. This indicates that the absorption was variable in time, but the absorption is initially not substantial and/or it reduced so quickly that it is unlikely a detection would have been possible.
 %Note that this is similar to simulations indicated by a blue circle, but here we get more time variability than in that case.
 \\
 \\
 
 %Blue Square & All criteria \emph{except} the criteria at $t=t_f$ are satisfied. This indicates there was a significant change in absorption over the time of the simulation, but there was too large of an absorption effect at the end of the simulation.
 
 %Blue Three Pointed Prong & The only failing criteria is the criteria on $\tau_{0.1-10}(0.1)$. This indicates we see strong absorption throughout the duration of the simulation (including at late times). This indicates nearly all of the radiation is being absorbed, leading to uninteresting results.

 \hline
 \hline
\end{tabular}
\end{table*}

Tables \ref{table:Sim_Results_LGRBs} and \ref{table:Sim_Results_SGRBs} as well as Figure \ref{fig:Significant_absorption_map} summarize our results with respect to which bursts satisfy our criteria for "potentially observable variable absorption". We see that these criteria are met for LGRBs only for the following combinations of $n_0$ and $H$ (presented as ordered pairs of the form $(n_0 \textrm{ [cm}^{-3}\textrm{]}, H \textrm{ [cm]})$):

$(10^8, 10^{16})$, $(10^7, 10^{16})$, $(10^7, 10^{17})$, $(10^6, 10^{17})$, $(10^6, 10^{18})$, $(10^5, 10^{17})$, $(10^5, 10^{18})$, $(10^4, 10^{18})$. 

For SGRBs, we see that the criteria are met only for the following combinations of $n_0$ and $H$:

$(10^{10}, 10^{14})$, $(10^9, 10^{14})$, $(10^9, 10^{15})$, $(10^8, 10^{15})$, $(10^8, 10^{16})$, $(10^7, 10^{15})$, $(10^7, 10^{16})$, $(10^7, 10^{17})$, $(10^6, 10^{16})$, $(10^6, 10^{17})$, $(10^5, 10^{16})$, $(10^5, 10^{17})$, $(10^5, 10^{18})$, $(10^4, 10^{17})$, $(10^4, 10^{18})$.

Outside of this range, we can intuitively understand the simulation failing our criteria for one of the two reasons below:

\begin{itemize}
    \item[1.] The medium is not dense or extended enough, and the optical depth remains low for the entire duration of the simulation. The radiation is then passing through the medium without ever being significantly absorbed.
    \item[2.] The medium is very dense and/or very extended, causing the optical depth to be large throughout the duration of the simulation. Thus, nearly all the radiation is being absorbed and there is no dynamical feedback between the medium and the radiation field.
\end{itemize}

While a larger and more refined grid search in the $(n_0, H)$ parameter space is needed to make a conclusive statement about where the potentially observable and variable absorption occurs, we make the following two conclusions based on our results: 
\begin{itemize}
    \item[1.] Variability in absorption is potentially observable for long GRBs emitted within dense environments when 
    \begin{equation}
    10^{4} \left(\frac{H}{10^{17}\textrm{ cm}}\right)^{-3} \leq n_0 \leq 10^{9} \left(\frac{H}{10^{17}\textrm{ cm}}\right)^{-3}.
    \label{eqn:bounds_on_interesting_LGRBs}
\end{equation}    
    
    \item[2.] Variability in absorption is potentially observable for short GRBs emitted within dense environments when
    
    \begin{equation}
    10^{3} \left(\frac{H}{10^{17}\textrm{ cm}}\right)^{-2} \leq n_0 \leq 10^{7} \left(\frac{H}{10^{17}\textrm{ cm}}\right)^{-2}.
    \label{eqn:bounds_on_interesting_SGRBs}
    \end{equation}
    
\end{itemize}

It is worth emphasizing here that these simulations and the results obtained, even though aimed at exploring a specific disk environment, have broader applicability. Results in these equations and Figure~\ref{fig:Significant_absorption_map} are applicable to any absorbing cloud with the density and spatial scale reported, as long as the gas is not hot enough to be fully ionized. %These might be, for example, a disk from a low-luminosity AGN, or even a molecular cloud surrounding the burst progenitor. 
Additionally, while in the following section we 
link these results to the specific structure of AGN accretion disks which are geometrically thin for most of their radial extent, we note that our simulations are rather independent of the specific geometry of the absorbing region, since they are line-of-sight calculations.

\subsection{Implications for AGN Disks}

\begin{figure}
     \centering
     \includegraphics[width=\columnwidth]{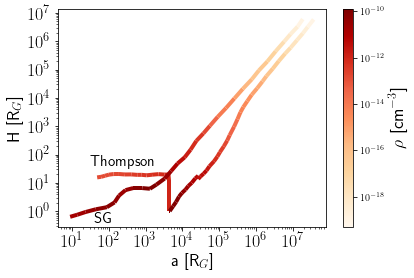}
     \caption{The central mass density and scale height of the Thompson and Sirko/Goodman (SG) AGN disk models as a function of radial position in the disk (in units of gravitational radii). The central mass density refers to the density in the plane of the disk, while the scale height encodes information about how quickly the density falls of as we move out of the plane of the disk.}
     \label{fig:AGN_models}
\end{figure}

When mapping conditions on $n_0$ and $H$ into conditions on location in an AGN disk and SMBH mass, one must pick a particular AGN accretion disk model to use. There are multiple reasonable choices here, such as the Shakura-Sunyaev disk model \citep{ShakuraSunyaev1973}, the Sirko \& Goodman (SG) model \citep{Sirko2003}, or the Thompson (TQM) model \citep{Thompson2005}. Both the SG model and the TQM model are improvements on the Shakura-Sunyaev model in that they are specific to AGN disks (as opposed to a general accretion disk). In particular, the SG model is thought to be a more accurate model of inner AGN disks, while the TQM model is thought to be better at describing the outer parts of those disks \citep{Fabj2020}. We thus map our conditions in the $(n_0, H)$ - space into conditions on location and SMBH mass for the TQM and SG models separately.

Figure \ref{fig:AGN_models} shows the density and scale height profiles for both of the AGN disk models we are considering. The density shown in the figure is the density in the plane of the disk (the central density).  
 
Using the AGN models, we can then find the radial location in the disk that this $\rho_0=n_0 m_p$ (where $m_p$ is the proton mass) corresponds to. After finding this, we can use what we know about the scale height in the AGN model to map the scale height to SMBH mass using the relation $R_G = G M / c^2$.

\begin{figure}
     \centering
     \begin{subfigure}[b]{\columnwidth}
         \centering
         \includegraphics[width=\columnwidth]{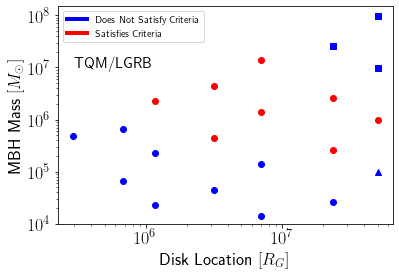}
     \end{subfigure}
     \hfill
     \begin{subfigure}[b]{\columnwidth}
         \centering
         \includegraphics[width=\columnwidth]{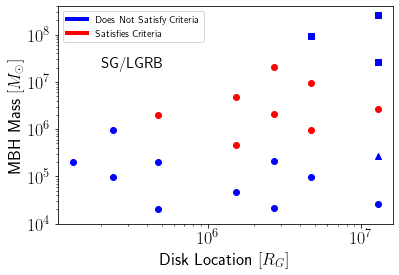}
     \end{subfigure}
     \caption{A parameter space mapping of where we see potentially observable variable absorption for LGRBs in both the TQM (top) and SG (bottom) AGN models. The red dots indicate simulations where our two criteria for absorption are met, while the blue dots indicate the opposite (meanings of the different shapes are given in Table \ref{table:marker_meanings}). Simulations where the mass of the black hole maps to a value less than $10^4$ $M_{\odot}$ are excluded since black holes of this mass, when accreting, cannot be classified as AGN.
     }
     \label{fig:AGN_parameter_space_maps_LGRBs}
\end{figure}

\begin{figure}
     \centering
     \begin{subfigure}[b]{\columnwidth}
         \centering
         \includegraphics[width=\columnwidth]{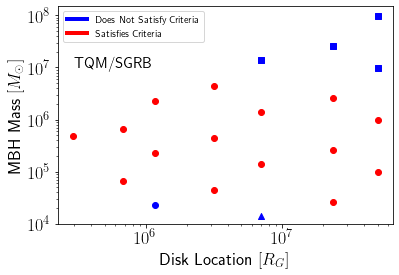}
     \end{subfigure}
     \hfill
     \begin{subfigure}[b]{\columnwidth}
         \centering
         \includegraphics[width=\columnwidth]{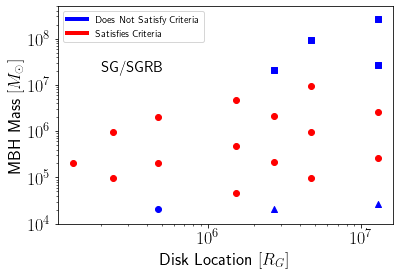}
     \end{subfigure}
     \caption{Same as in Fig.~\ref{fig:AGN_parameter_space_maps_LGRBs} but for SGRBs.
     }
     \label{fig:AGN_parameter_space_maps_SGRBs}
\end{figure}

With the method described, we can easily create figures that are complementary to Figure \ref{fig:Significant_absorption_map}, where instead of showing the simulations in the $(n_0, H)$ parameter space, we show the simulations in the $($disk location$,$ SMBH mass$)$ parameter space. Figures \ref{fig:AGN_parameter_space_maps_LGRBs} and \ref{fig:AGN_parameter_space_maps_SGRBs} show this for both the TQM and SG AGN models. Note that the regions with potentially observable variable absorption in Figures \ref{fig:AGN_parameter_space_maps_LGRBs} and \ref{fig:AGN_parameter_space_maps_SGRBs}  follow a more complicated pattern than 
in Figure~\ref{fig:Significant_absorption_map}
as a result of the non-monotonic density and scale height profiles of the SG and TQM AGN disk models (cfr Fig.~\ref{fig:AGN_models}).
Before commenting on the specific results  from those disk models, we wish to emphasize that, for our study, we have used the theoretical disk profiles  up to the outer radii often considered in the literature on stars in AGN disks (e.g. \citealt{Fabj2020}). However, the precise location at which the outer disk effectively cuts off is rather uncertain. This makes the possibility of observing a GRB with variable absorption exciting because it would allow us to put independent observational constraints on the outer disk regions.

Based on inspection of Figures \ref{fig:AGN_parameter_space_maps_LGRBs} and \ref{fig:AGN_parameter_space_maps_SGRBs}, we make the following conclusions that act in a complementary way to the conclusions made in section \ref{sec:Burst_Classification} but are specific for AGN disks: 

\begin{itemize}
    \item Variability in absorption is potentially observable for LGRBs emitted within dense environments only when the mass of the MBH falls within a band between $10^5 M_{\odot}$ and $5\times 10^7 M_{\odot}$.
    \item Variability in absorption is potentially observable for SGRBs emitted within dense environments only when the mass of the MBH falls within a band between $10^4 M_{\odot}$ and $10^7 M_{\odot}$.
\end{itemize}

We note that, while typical AGNs are found to have MBHs larger than $\sim 10^6M_\odot$, for the analysis reported above we have formally extended the lower mass limit  of the MBH to below this  value. BHs in the mass range of $10^4-10^6M_\odot$ would be found in dwarf galaxy-AGN systems, of which a few examples have been discovered in recent years, down to a $50,0000~M_\odot$  case \citep{Baldassare2015}. The occurrence of transients in the disks of these systems, while less likely due to their smaller size,  would allow one to probe their structure, since time-variable absorption of the transient spectra would be especially enhanced in these disks.

\begin{figure*}
    \centering
    \includegraphics[width=1.0\textwidth]{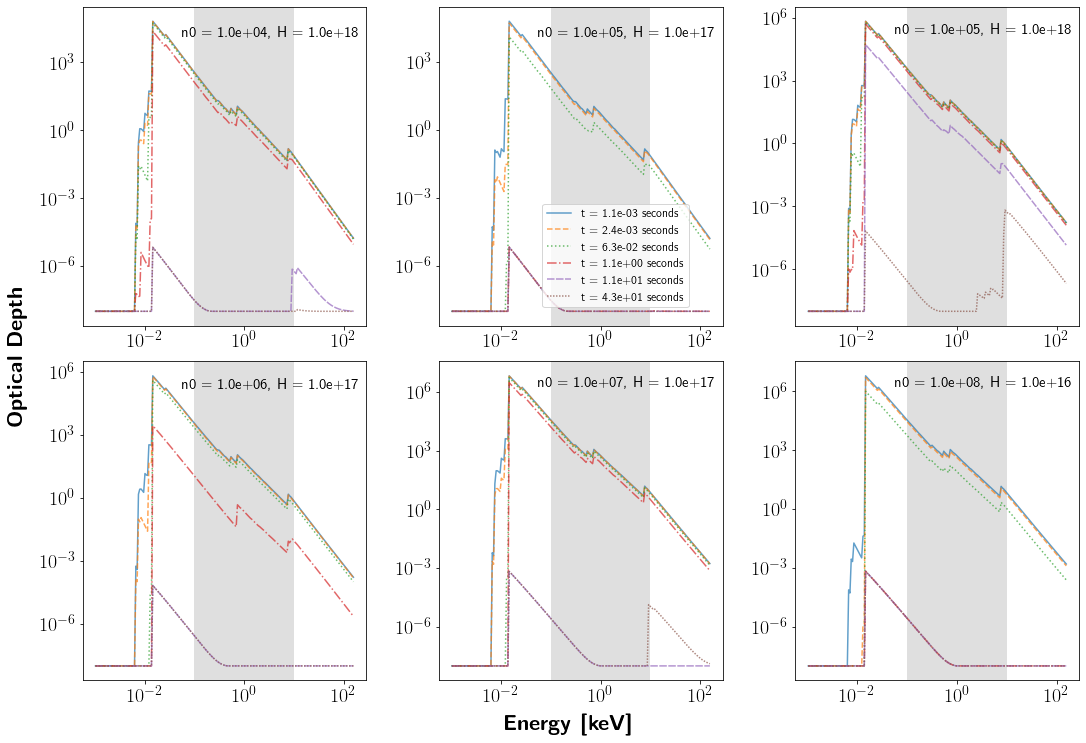}
    \centering
    \caption{Optical depth of a selection of LGRB simulations that meet our criteria for potentially observable absorption. Shown here is the optical depth as a function of frequency at various times during the simulations. The shaded region indicates the band of energies between 0.1keV and 10keV (the region used to define the absorption criteria). Each plot shows the optical depth at six separate times. For a view of optical depths for every LGRB simulation run, see fig.\ref{fig:optical_depth_grid_LGRBs}.}
    \label{fig:interesting_OD_LGRBs}
\end{figure*}

\begin{figure*}
    \centering
    \includegraphics[width=1.0\textwidth]{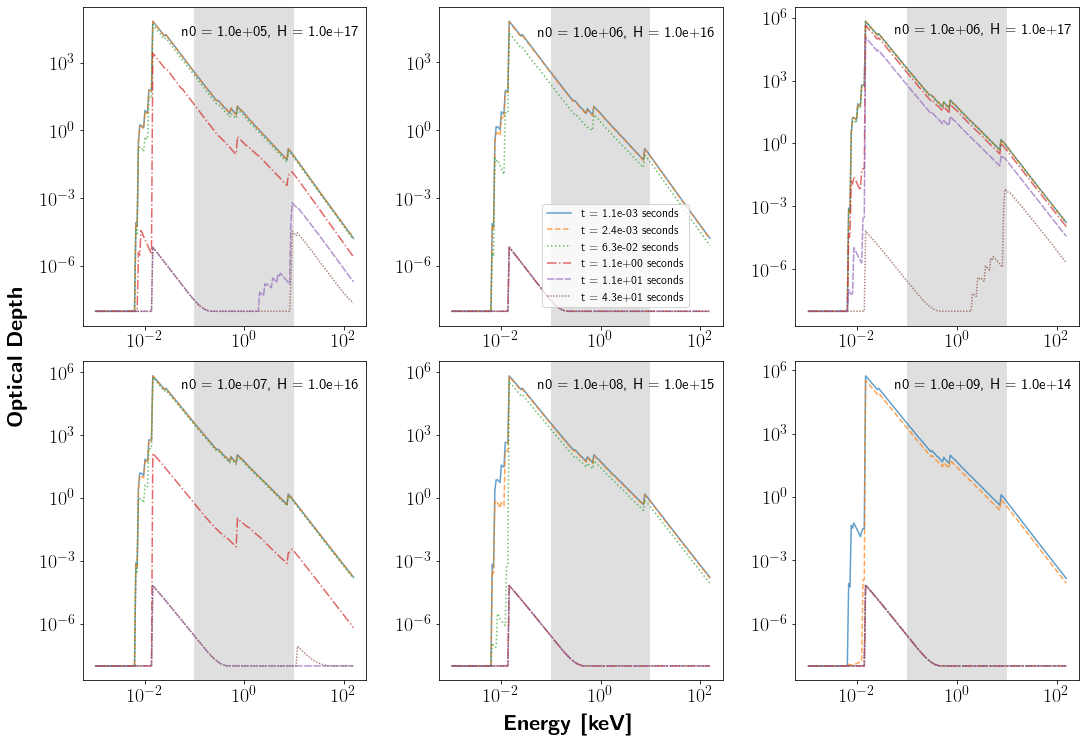}
    \centering
    \caption{Same as fig.\ref{fig:interesting_OD_LGRBs} but for SGRBs. For a view of optical depths for every SGRB simulation run, see fig.\ref{fig:optical_depth_grid_SGRBs}.}
    \label{fig:interesting_OD_SGRBs}
\end{figure*}

In order to better connect to the observables, and that is the time-dependent spectra of the transients, we begin by investigating the energy-dependent behaviour of the opacity, while the medium gets photoionized by the radiation from the transient. This is shown
for selected grid points of our ($n_0,H$) grid in Figs.~\ref{fig:interesting_OD_LGRBs} and \ref{fig:interesting_OD_SGRBs} for the cases of LGRBs and SGRBs, respectively.
 We show the same plots, but including every point of our $(n_0,H)$ study grid, in Figs.~\ref{fig:optical_depth_grid_LGRBs} and \ref{fig:optical_depth_grid_SGRBs} for the cases of LGRBs and SGRBs, respectively. Recall that the grid is limited by the condition
 $H n_0 \leq 10^{24}$ cm$^{-2}$ required to ensure transparency of the prompt $\gamma$-rays to Thompson scattering. Hence the panels which do not satisfy this condition have been omitted.

For each $(n_0,H)$ combination, we show the opacity at six times after the burst onset: $t = 0.0011~\textrm{ sec},\; t = 0.0024~\textrm{ sec}, \; t = 0.063~\textrm{ sec}, \; t = 1.1 \textrm{ sec}, \; t = 11~\textrm{ sec},\; \textrm{and } t = 43 \textrm{ sec}$. The times are chosen such that we can see the optical depth during important times in the burst's lifetime.
The general trend that we can infer from the figures is that of a more rapid time variability (signaling quick photoionization of the medium) for smaller medium densities and shorter scale heights (hence the region in the left bottom panels of the figures). Additionally, for the same medium parameters (hence corresponding panels between 
Figs.~\ref{fig:optical_depth_grid_LGRBs} and \ref{fig:optical_depth_grid_SGRBs}),
the most intense flux from LGRBs induces a quicker reduction of the opacity, as expected.  

\begin{figure*}
    \centering
    \includegraphics[width=1.0\textwidth]{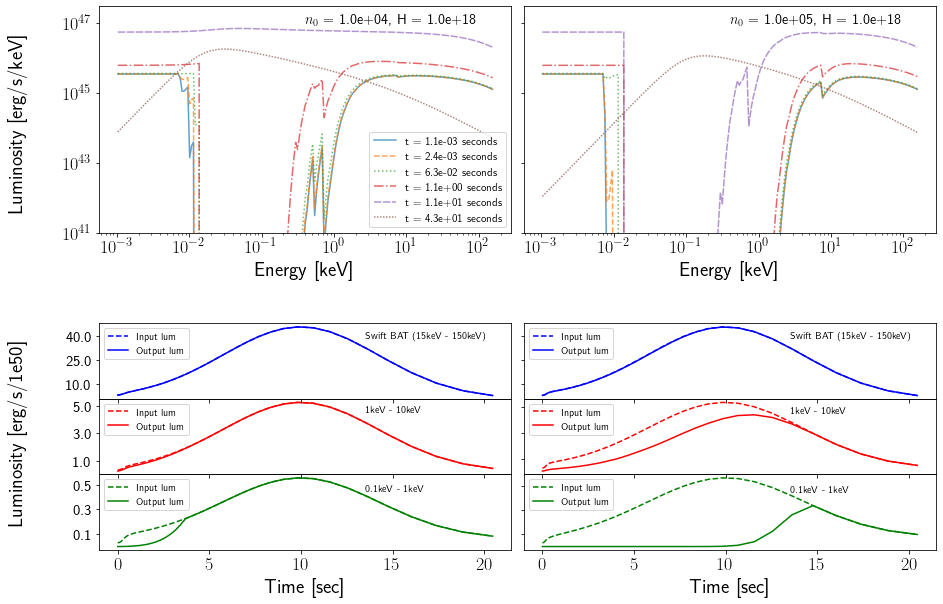}
    \centering
    \caption{In the top two panels, we show emergent luminosity spectra of LGRBs for two combinations of medium density and scale height, chosen among the cases displaying large optical depth variations during the early times of the transient. 
    As indicated in the label in the top left panel, each plot shows the luminosity at six separate times.
    In the bottom two panels, we show the emergent luminosity integrated over three different bands of energy.}
    \label{fig:emergent_lum_LGRBs}
\end{figure*}

\begin{figure*}
    \centering
    \includegraphics[width=1.0\textwidth]{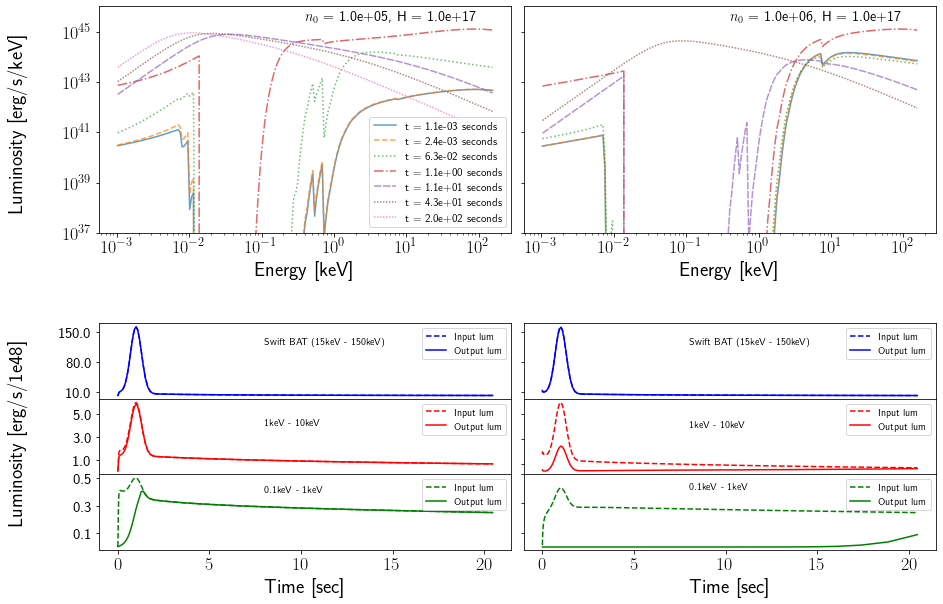}
    \centering
    \caption{Same as Fig.\ref{fig:emergent_lum_LGRBs} but for SGRBs.}
    \label{fig:emergent_lum_SGRBs}
\end{figure*}

From a closer inspection of Figs.\ref{fig:interesting_OD_LGRBs}, \ref{fig:interesting_OD_SGRBs},~\ref{fig:optical_depth_grid_LGRBs}, and \ref{fig:optical_depth_grid_SGRBs} it is evident that there are situations, i.e. combinations of $(n_0,H)$, for which the opacity varies considerably from the UV/soft X-rays to the hard X-rays. 
This variability results in the appearance of transients with unusual spectral properties and evolution, as shown in Figs.~\ref{fig:emergent_lum_LGRBs} and \ref{fig:emergent_lum_SGRBs} for LGRBs and SGRBs, respectively.
Early-time UV and soft X-ray emission would be suppressed, only to rapidly emerge later with a rebrightening that proceeds from the harder to the softer radiation down to the UV. 
As shown in the bottom panels of Figures~\ref{fig:emergent_lum_LGRBs} and~\ref{fig:emergent_lum_SGRBs}, the identification of such transients would be better obtained by comparing soft and hard X-ray light curves. Care would be needed as gamma-ray bursts have, at least in some cases, intrinsic hard to soft spectral evolution, which could mimic the effect described here. Spectra from absorbed bursts would, however, appear extremely X-ray poor. In addition, their spectral shape would be consistent with a power-law spectrum absorbed at the source. Finally, the hard to soft evolution should be strictly monotonic, since the recombination time of free electrons onto ions is longer --- even at high density --- than the few seconds to tens of seconds considered in this paper.
We further note that any time variability in the optical band due to dust destruction \citep{Waxman2000} would occur on a much too short timescale to be detectable, since in very dense regions the timescale for dust destruction is faster than that for gas photoionization \citep{Perna2003}. 

Before concluding, we need to remind that observability of transients from AGN disks is clearly dependent on their brightness exceeding that of the AGN disk itself. Most AGNs have luminosities in the $10^{43}-10^{47}$~erg~s$^{-1}$ range distributed across a large spectrum but with a large fraction in the UV and optical bands, and no apparent strong correlation with the SMBH mass \citep{Woo2002}. In the X-rays, the luminosity function of AGNs is characterized by a power-law \citep{Gilli2007}. Measurements of the 2-10~keV luminosity function in the low-redshift Universe \citep{Ueda2003} show that low-luminosity sources (with luminosities $\sim 10^{41}-10^{42}$~erg~s$^{-1}$)  outnumber the higher luminosity ones with power $\sim 10^{46}-10^{47}$~erg~s$^{-1}$ by about 7 orders of magnitude. Hence, comparing with Figs.~~\ref{fig:emergent_lum_LGRBs} and \ref{fig:emergent_lum_SGRBs}, we can conclude that the time-variable higher energy component of the GRB transients, especially in the X-rays, is expected to be detectable at high signal-to noise for the majority of AGN disks, and especially for LGRBs. This would be even more so in the subclass of 'low-luminosity' AGNs, which have bolometric luminosities around $10^{39}-10^{41}$~erg~s$^{-1}$ \citep{Maoz2007}.

%-------------------------------------------------------
%Plots of optical depth for significantly absorbed GRBs
%-------------------------------------------------------

%_____________________________
%  GRID FIGURES
%_____________________________

%_____________________
% TABLE OF TAU VALUES AND WHETHER OR NOT SIG ABSORPTION OCCURRED FOR LONG GRBS
%_____________________

\section{Conclusions \& Discussion}

In this work, we presented a grid of simulations investigating the absorption of gamma-ray bursts emitted from within dense environments. We presented a  definition of what "potentially observable variable absorption"  means in this context and then proceeded to present which of our simulations met this definition. Our results led us to conclude that potentially observable variable absorption of LGRBs in dense environments only occurs in a certain band of $(n_0, H)$ parameter space, in particular, only when $10^{5} (\frac{H}{10^{17}\textrm{ cm}})^{-3} \leq n_0  \leq 10^{8} (\frac{H}{10^{17}\textrm{cm}})^{-3}$. We made an analogous conclusion for SGRBs, namely that potentially observable variable absorption only occurs when $10^{3} (\frac{H}{10^{17}\textrm{ cm}})^{-2} \leq n_0 \leq 10^{6} (\frac{H}{10^{17}\textrm{cm}})^{-2}$. We then transformed our findings in the $(n_0, H)$ parameter space to findings in the $($disk location$,$ SMBH mass$)$ parameter space by choosing two relevant AGN disk models. Here we found that for both models, potentially observable variable absorption seems to only occur in a narrow band of SMBH masses and for locations at a significant distance from the central SMBH. For LGRBs, this band is characterized by SMBH masses between $10^5 M_{\odot}$ and $10^7 M_{\odot}$. For SGRBs, the band is characterized by SMBH masses between $10^4 M_{\odot}$ and $10^7 M_{\odot}$. 

Independently on the specific disk structure considered, it is clear that bursts exploding in regions with large density and size ranging from a fraction to tens of parsecs are significantly affected by photon propagation. These transients are initially highly absorbed and would be completely dark in the UV and soft X-ray bands ($h\nu\lesssim 10$~keV) for a few seconds (Figures~\ref{fig:emergent_lum_LGRBs}, \ref{fig:emergent_lum_SGRBs},
\ref{fig:optical_depth_grid_LGRBs}, and~\ref{fig:optical_depth_grid_SGRBs}). Strong spectral evolution in the same bands is expected during the brightest pulses of the prompt emission, when the ionizing flux is higher. As time progresses, soft X-rays would emerge first, possibly followed by prompt UV emission. We have shown that the identification of such transients would be better obtained by comparing soft and hard X-ray light curves. 
 Additional confirmation that one is observing an initially absorbed burst would come from other indicators, such as a radio dark, or even optically dark burst \citep{Wang2022}, characteristics of a high-density environment. Alternatively, one may detect the telltale FRED pulse shape with overlaid variability that indicates a burst with a very close-by external shock onset \citep{Lazzati2022}. Finally, if a precise location is obtained from afterglow observations, a burst location coincident with the center of the host galaxy would be a telltale sign of a candidate burst for further spectral study. On the other hand, the information encoded in the time evolution of the variability, once established, would offer unique insight on the properties of the immediate burst surrounding medium.

While our findings have focused on specific AGN models, the simulations  only assume that the density of the medium along the line-of-sight falls off as a Gaussian. Thus, our findings in the $(n_0, H)$ parameter space apply to any environment where the density falls off in this manner. Our results have therefore the potential to be used as probes of the disk structure, which is still debated and can vary significantly among models. Should a GRB be detected within an accretion disk, its properties and spectral evolution could be used to probe the local density structure of the disk.

\section*{Acknowledgements}

RP and MR acknowledge support by NSF award AST-2006839. DL acknowledges support from NSF award AST-1907955.

\section*{Data Availability}

All data needed to reproduce this work can be found on Zenodo at the following URL: \newline \href{https://doi.org/10.5281/zenodo.7267661}{https://doi.org/10.5281/zenodo.7267661}

%%%%%%%%%%%%%%%%%%%% REFERENCES %%%%%%%%%%%%%%%%%%

% The best way to enter references is to use BibTeX:

\bibliographystyle{mnras}
\bibliography{references} % if your bibtex file is called example.bib

% Alternatively you could enter them by hand, like this:
% This method is tedious and prone to error if you have lots of references
%\begin{thebibliography}{99}
%\bibitem[\protect\citeauthoryear{Author}{2012}]{Author2012}
%Author A.~N., 2013, Journal of Improbable Astronomy, 1, 1
%\bibitem[\protect\citeauthoryear{Others}{2013}]{Others2013}
%Others S., 2012, Journal of Interesting Stuff, 17, 198
%\end{thebibliography}

%%%%%%%%%%%%%%%%%%%%%%%%%%%%%%%%%%%%%%%%%%%%%%%%%%

%%%%%%%%%%%%%%%%% APPENDICES %%%%%%%%%%%%%%%%%%%%%

\appendix

\section{All Optical Depth Plots}

Here we present grid plots showing optical depth as a function of energy at six different times for both long and short GRBs, for every combination of $n_0$ and $H$. These figures are an extension of figures \ref{fig:interesting_OD_LGRBs} and \ref{fig:interesting_OD_SGRBs}, which only show a subset of the simulations. Figures marked with an asterisk in these grids are the simulations that meet our criteria for having potentially observable variable absorption.

\begin{figure*}
    \centering
    \includegraphics[width=1.0\textwidth]{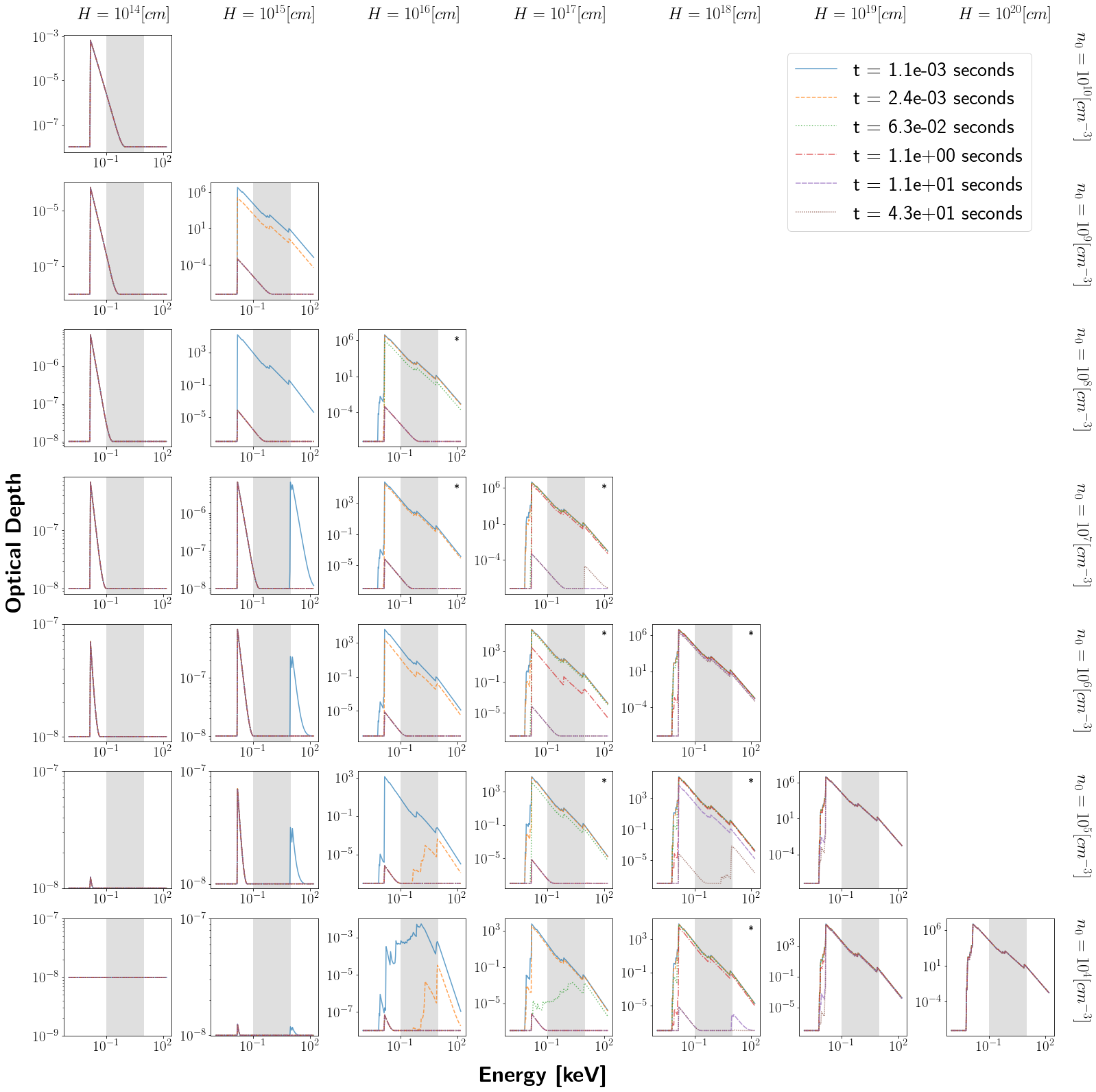}

    \centering
     \caption{Optical depth of LGRB simulations as a function of frequency at various times during the simulation,
     for the $(n_0,H)$ grid of our study.
      The grid is laid out such that $H$ increases to the right in a logarithmic fashion from $10^{14}$ cm to $10^{20}$ cm and $n_0$ increases vertically up the grid from $10^4$ cm$^{-3}$ to $10^{10}$ cm$^{-3}$. The shaded region indicates the band of energies between 0.1 keV and 10 keV (the region used to define potentially observable variable absorption) and plots with an asterisk correspond to simulations that satisfy our criteria of potentially observable variable absorption. 
      Missing graphs in the grid here represent absorbing media with column density $H n_0 \geq 10^{24}$ cm$^{-2}$, which we omit from our analysis.
     Each plot shows optical depth at six separate times.
     We see that the optical depth (and hence overall absorption) is more rapidly reduced at both smaller medium densities and shorter scale heights.}
     \label{fig:optical_depth_grid_LGRBs}
\end{figure*}

\begin{figure*}
    \centering
    \includegraphics[width=1.0\textwidth]{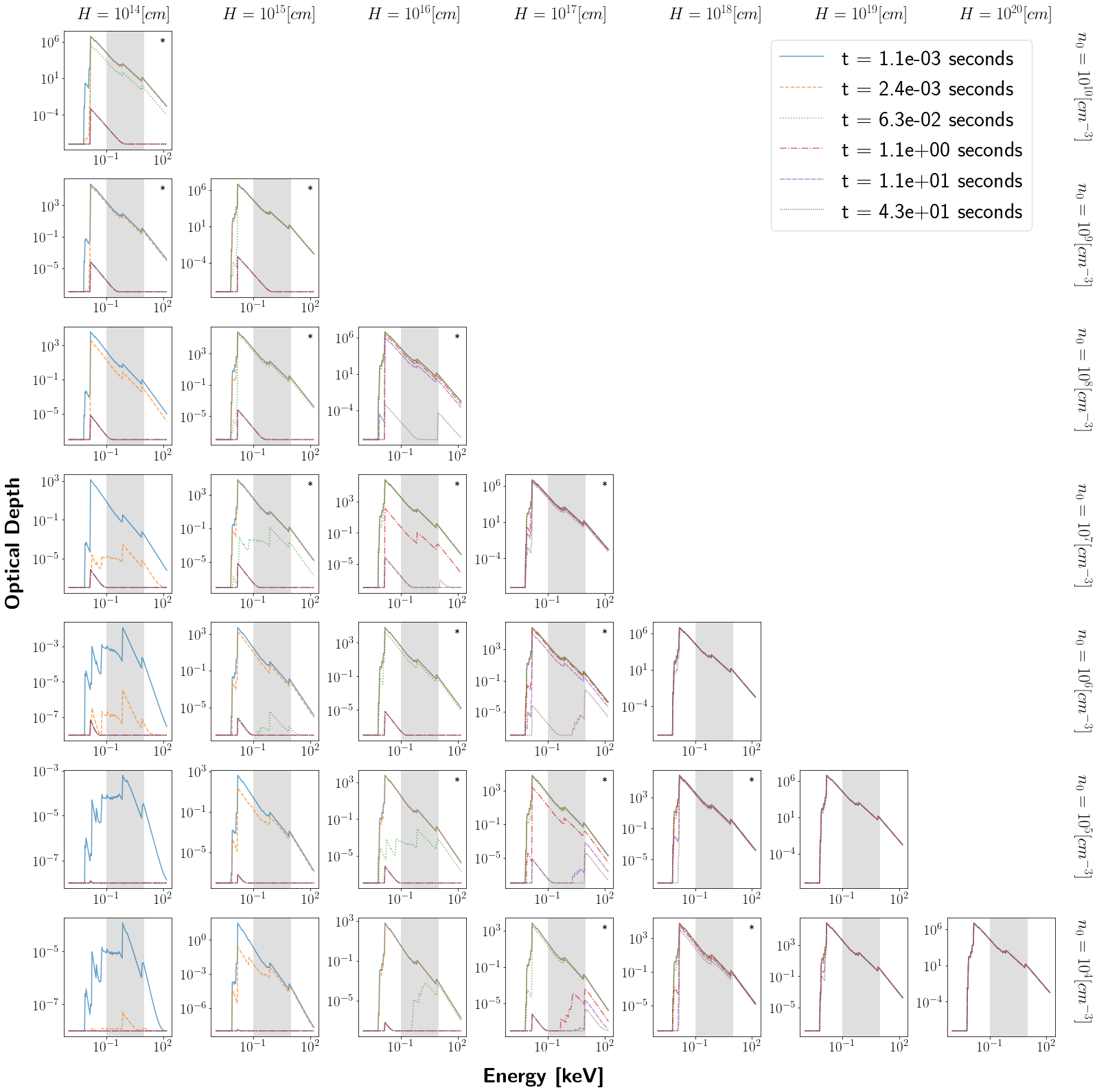}

    \centering
     \caption{Same as Fig.~\ref{fig:optical_depth_grid_LGRBs} but for SGRBs.}

     \label{fig:optical_depth_grid_SGRBs}
\end{figure*}

\section{Optical Depth Data}

%_____________________
% TABLE OF TAU VALUES AND WHETHER OR NOT SIG ABSORPTION OCCURRED FOR LONG GRBS
%_____________________

\maketitle % <==========================================================
\begin{table*} % <======================================================
  \caption{We present the average optical depth in the 0.1 - 10 keV range at 0.1 seconds for each of the LGRB simulations. Also shown is the ratio of average optical depth at 0.1 seconds to the average optical depth at $t_{f}$. An additional column is added to indicate whether the simulation satisfies both of our criteria for potentially observable variable absorption to have occurred (the precise definition of this is given in section \ref{sec:Simulation_Description_and_Results}).}\label{table:Sim_Results_LGRBs} % <========================
\begin{tabular}{ |p{4cm}|p{4cm}|p{4cm}|p{4cm}| }
\textbf{Long GRBs}
\\
\\
 \hline
 \hline
 (n$_0$ [cm$^{-2}$], H [cm])& 
 %$\tau_{0.1-10}(t_f)$& 
 $\tau_{0.1-10}(0.1 sec)$& $\frac{\tau_{0.1-10}(0.1 sec)}{\tau_{0.1-10}(t_f)}$& Criteria Satisfied?\\
 \hline
 $(1.0e+04, 1.0e+14)$   & $1.00e-08$    & $1.00e+00$    & 
False
  \\ 
 \\

$(1.0e+04, 1.0e+15)$   & $1.00e-08$    & $1.00e+00$    & 
False
  \\ 
 \\

$(1.0e+04, 1.0e+16)$   & $1.00e-08$    & $1.00e+00$    & 
False
  \\ 
 \\

$(1.0e+04, 1.0e+17)$   & $2.40e-01$    & $2.40e+07$    & 
False
  \\ 
 \\

$(1.0e+04, 1.0e+18)$   & $1.50e+01$    & $1.48e+09$    & 
True
  \\ 
 \\

$(1.0e+04, 1.0e+19)$   & $1.65e+02$    & $1.86e+00$    & 
False
  \\ 
 \\

$(1.0e+04, 1.0e+20)$   & $1.67e+03$    & $1.06e+00$    & 
False
  \\ 
 \\

$(1.0e+05, 1.0e+14)$   & $1.00e-08$    & $1.00e+00$    & 
False
  \\ 
 \\

$(1.0e+05, 1.0e+15)$   & $1.00e-08$    & $1.00e+00$    & 
False
  \\ 
 \\

$(1.0e+05, 1.0e+16)$   & $1.00e-08$    & $1.00e+00$    & 
False
  \\ 
 \\

$(1.0e+05, 1.0e+17)$   & $8.82e+00$    & $8.73e+08$    & 
True
  \\ 
 \\

$(1.0e+05, 1.0e+18)$   & $1.58e+02$    & $9.77e+09$    & 
True
  \\ 
 \\

$(1.0e+05, 1.0e+19)$   & $1.66e+03$    & $1.25e+00$    & 
False
  \\ 
 \\

$(1.0e+06, 1.0e+14)$   & $1.00e-08$    & $1.00e+00$    & 
False
  \\ 
 \\

$(1.0e+06, 1.0e+15)$   & $1.00e-08$    & $1.00e+00$    & 
False
  \\ 
 \\

$(1.0e+06, 1.0e+16)$   & $1.01e-08$    & $1.00e+00$    & 
False
  \\ 
 \\

$(1.0e+06, 1.0e+17)$   & $1.28e+02$    & $7.91e+09$    & 
True
  \\ 
 \\

$(1.0e+06, 1.0e+18)$   & $1.63e+03$    & $4.08e+00$    & 
True
  \\ 
 \\

$(1.0e+07, 1.0e+14)$   & $1.00e-08$    & $1.00e+00$    & 
False
  \\ 
 \\

$(1.0e+07, 1.0e+15)$   & $1.01e-08$    & $1.00e+00$    & 
False
  \\ 
 \\

$(1.0e+07, 1.0e+16)$   & $1.40e+01$    & $8.66e+08$    & 
True
  \\ 
 \\

$(1.0e+07, 1.0e+17)$   & $1.48e+03$    & $1.61e+10$    & 
True
  \\ 
 \\

$(1.0e+08, 1.0e+14)$   & $1.01e-08$    & $1.00e+00$    & 
False
  \\ 
 \\

$(1.0e+08, 1.0e+15)$   & $1.62e-08$    & $1.00e+00$    & 
False
  \\ 
 \\

$(1.0e+08, 1.0e+16)$   & $8.42e+02$    & $9.16e+09$    & 
True
  \\ 
 \\

$(1.0e+09, 1.0e+14)$   & $1.62e-08$    & $1.00e+00$    & 
False
  \\ 
 \\

$(1.0e+09, 1.0e+15)$   & $9.20e-08$    & $1.00e+00$    & 
False
  \\ 
 \\

$(1.0e+10, 1.0e+14)$   & $9.20e-08$    & $1.00e+00$    & 
False
  \\ 
 \\

 \hline
 \hline
 \end{tabular}
 \end{table*}
 
 \begin{table*} % <======================================================
  \caption{The same as table \ref{table:Sim_Results_LGRBs}, but this time presenting short GRB simulation data.}\label{table:Sim_Results_SGRBs} % <========================
\begin{tabular}{ |p{4cm}|p{4cm}|p{4cm}|p{4cm}| }
\textbf{Short GRBs}
\\
\\
 \hline
 \hline
 (n$_0$ [cm$^{-2}$], H [cm])& 
 %$\tau_{0.1-10}(t_f)$& 
 $\tau_{0.1-10}(0.1 sec)$& $\frac{\tau_{0.1-10}(0.1 sec)}{\tau_{0.1-10}(t_f)}$& Criteria Satisfied?\\
 \hline

$(1.0e+04, 1.0e+14)$   & $1.00e-08$    & $1.00e+00$    & 
False
  \\ 
 \\

$(1.0e+04, 1.0e+15)$   & $3.83e-05$    & $3.83e+03$    & 
False
  \\ 
 \\

$(1.0e+04, 1.0e+16)$   & $8.94e-02$    & $8.94e+06$    & 
False
  \\ 
 \\

$(1.0e+04, 1.0e+17)$   & $1.59e+00$    & $1.59e+08$    & 
True
  \\ 
 \\

$(1.0e+04, 1.0e+18)$   & $1.66e+01$    & $7.43e+02$    & 
True
  \\ 
 \\

$(1.0e+04, 1.0e+19)$   & $1.67e+02$    & $1.17e+00$    & 
False
  \\ 
 \\

$(1.0e+04, 1.0e+20)$   & $1.67e+03$    & $1.02e+00$    & 
False
  \\ 
 \\

$(1.0e+05, 1.0e+14)$   & $1.00e-08$    & $1.00e+00$    & 
False
  \\ 
 \\

$(1.0e+05, 1.0e+15)$   & $3.73e-06$    & $3.73e+02$    & 
False
  \\ 
 \\

$(1.0e+05, 1.0e+16)$   & $7.68e-01$    & $7.68e+07$    & 
True
  \\ 
 \\

$(1.0e+05, 1.0e+17)$   & $1.57e+01$    & $1.56e+09$    & 
True
  \\ 
 \\

$(1.0e+05, 1.0e+18)$   & $1.66e+02$    & $2.96e+00$    & 
True
  \\ 
 \\

$(1.0e+05, 1.0e+19)$   & $1.67e+03$    & $1.08e+00$    & 
False
  \\ 
 \\

$(1.0e+06, 1.0e+14)$   & $1.00e-08$    & $1.00e+00$    & 
False
  \\ 
 \\

$(1.0e+06, 1.0e+15)$   & $1.32e-02$    & $1.32e+06$    & 
False
  \\ 
 \\

$(1.0e+06, 1.0e+16)$   & $1.24e+01$    & $1.22e+09$    & 
True
  \\ 
 \\

$(1.0e+06, 1.0e+17)$   & $1.63e+02$    & $1.00e+10$    & 
True
  \\ 
 \\

$(1.0e+06, 1.0e+18)$   & $1.67e+03$    & $1.39e+00$    & 
False
  \\ 
 \\

$(1.0e+07, 1.0e+14)$   & $1.00e-08$    & $1.00e+00$    & 
False
  \\ 
 \\

$(1.0e+07, 1.0e+15)$   & $6.84e+00$    & $6.77e+08$    & 
True
  \\ 
 \\

$(1.0e+07, 1.0e+16)$   & $1.56e+02$    & $9.64e+09$    & 
True
  \\ 
 \\

$(1.0e+07, 1.0e+17)$   & $1.66e+03$    & $1.12e+01$    & 
True
  \\ 
 \\

$(1.0e+08, 1.0e+14)$   & $1.01e-08$    & $1.00e+00$    & 
False
  \\ 
 \\

$(1.0e+08, 1.0e+15)$   & $1.28e+02$    & $7.89e+09$    & 
True
  \\ 
 \\

$(1.0e+08, 1.0e+16)$   & $1.63e+03$    & $1.77e+10$    & 
True
  \\ 
 \\

$(1.0e+09, 1.0e+14)$   & $1.41e+01$    & $8.68e+08$    & 
True
  \\ 
 \\

$(1.0e+09, 1.0e+15)$   & $1.48e+03$    & $1.61e+10$    & 
True
  \\ 
 \\

$(1.0e+10, 1.0e+14)$   & $8.10e+02$    & $8.81e+09$    & 
True
  \\ 
 \\

 \hline
 \hline
 \\
 \\
\end{tabular}
\end{table*}

%%%%%%%%%%%%%%%%%%%%%%%%%%%%%%%%%%%%%%%%%%%%%%%%%%

%%%%%%%%%%%%%%%%%%%%%%%%%%%%%%%%%%%%%%%%%%%%%%%%%%

% Don't change these lines
\bsp	% typesetting comment
\label{lastpage}
\end{document}